\documentclass[letterpaper,12pt]{article}
\pdfoutput=1

\usepackage{jheppub}
\usepackage{amsmath}
\usepackage{graphicx}
\usepackage{subfig}
\usepackage{cancel}
\usepackage[dvipsnames]{xcolor}
\usepackage{color}
\usepackage{mathrsfs}

\usepackage{amssymb}
\usepackage{epstopdf}
\newpage

\title{Microcanonical Path Integrals and the Holography of small Black Hole Interiors}

\author{Donald Marolf}

\affiliation{Department of Physics, University of California, Santa Barbara, CA 93106, USA}
\emailAdd{marolf@physics.ucsb.edu}

\abstract{We use a microcanonical path integral closely related to that introduced by Brown and York in 1992 to add new entries to the AdS/CFT dictionary concerning the interiors of small black holes.  Stationary points of such path integrals are also stationary points of more standard canonical-type path integrals with fixed boundary metric, but the condition for dominance is now maximizing Hubeny-Rangamani-Takayanagi entropy at fixed energy.  As a result, such path integrals can bring to the fore saddles that fail to dominate in more familiar contexts.  We use this feature to argue that the standard Kruskal-like two-sided extension of small AdS black holes with energy $E_0$ is dual to a microcanonical version of the thermofield double state for AdS black holes that maximize the microcanonical bulk entropy at this energy.  We also comment on entanglement in such states and on quantum effects that become large when the energy-width of the microcanonical ensemble is sufficiently small.
}

\begin{document}
\maketitle

\section{Introduction}
\label{sec:Introduction}

Developing a full understanding of the bulk-to-boundary dictionary remains a long-standing goal in the study of Anti-de Sitter/Conformal Field Theory duality (AdS/CFT).   While much of the recent activity in this regard (e.g. \cite{Jafferis:2015del,Botta-Cantcheff:2015sav,Dong:2016eik,Christodoulou:2016nej,Faulkner:2017vdd,Marolf:2017kvq}) has focussed on bulk quantum fluctuations (and thus on effects suppressed by powers of the bulk Newton constant $G$), our goal below is to study issues at {\it leading} order
in the bulk Newton's constant $G$ associated with the interiors of eternal black holes.

In particular, while Maldacena's well-known Euclidean path-integral argument \cite{Maldacena:2001kr} relating thermofield-double (TFD) states in the dual quantum field theory (QFT) to bulk two-sided AdS-Schwarzschild black holes (figure 1) is easily generalized to more complicated manifolds (see \cite{Maldacena:2001kr} for brief comments, but also \cite{Skenderis:2009ju,Balasubramanian:2014hda,Marolf:2015vma,Maxfield:2016mwh,Marolf:2017shp,Marolf:2017vsk,Fu:2018kcp}), in the classical bulk limit it provides entries in this dictionary only for saddle points that dominate the bulk computation.  As a result, it relates TFD states to to standard two-sided AdS-Schwarzschild black holes only at energies above the Hawking-Page transition \cite{Hawking:1982dh}.  What then is the status of similar black holes below this threshold? We emphasize again that our concern lies with the interior structure of such black holes, and in particular with the Einstein-Rosen-like bridge connecting the two asymptotic regions.  In contrast, it is clear that small black holes formed by collapse of matter or by the Hawking evaporation of larger black holes are described by corresponding states in the dual QFT, see e.g. \cite{Asplund:2008xd,Hanada:2016pwv,Yaffe:2017axl,Berenstein:2018lrm} for arguments addressing the form of such QFT states.

\begin{figure}[t]
        \begin{center}
                \includegraphics[width=0.25\textwidth]{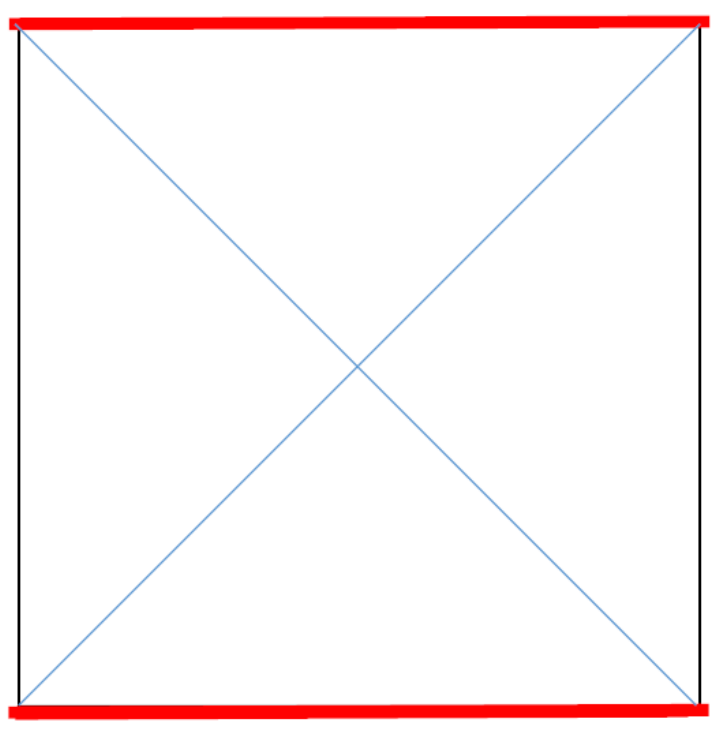}
        \end{center}
        \caption{A rough conformal diagram of a 2-sided AdS-Schwarzschild black hole.  The vertical (black) lines are the two disconnected components of the asymptotically-AdS conformal boundary.  The heavy horizontal (red) lines are the singularity.  In an exact conformal diagram these lines would not meet the AdS boundaries orthogonally (see explicit discussion in \cite{Fidkowski:2003nf} and implicit discussion surrounding figure 7 of \cite{Klosch:1995qv}).  The diagonal (blue) lines are the horizons.  The associated wormhole is almost traversable in the sense that the past event horizon of one boundary coincides with the future event horizon of the other, so that an infinitesimal perturbation of the geometry (violating the averaged null energy condition as in \cite{Gao:2016bin}) could render the wormhole traversable. }
        \label{fig:wormhole}
\end{figure}

We study small two-sided black holes below by considering new classes of path integrals associated with microcanonical relatives of TFD states.  Indeed, for window functions $f$ that are sharply peaked at $f(0)=1$ and for energies $E_0$ where black holes dominate the density of states, we will argue that the micro-canonical thermofield-double (MCTFD) state

\begin{equation}
\label{eq:MCTFD}
|\psi \rangle = \sum_E e^{-\beta E /2}f(E-E_0) |E\rangle |E\rangle
\end{equation}
on two copies of the dual QFT admits a small $G$ bulk description as the standard two-sided Kruskal-like extention of the entropically-dominant black hole at energy $E_0$. Here we assume a context with time-translation invariance, so that the two copies of the QFT (called left and right) have equivalent Hamiltonians $H_L, H_R$. In particular, as expected from \eqref{eq:MCTFD} the bulk solutions will be invariant under a Killing symmetry associated with the generator $H_L-H_R$ such that the two sides of the bulk are connected by an Einstein-Rosen-like ``almost-traversable wormhole."  By this we mean that, as in figure 1, the past horizon of one asymptotic AdS boundary coincides with the future horizon of the other.  We also discuss corrections at sub-leading orders in $G$. We have chosen to write the state \eqref{eq:MCTFD} in a form where it is not properly normalized, but instead has norm  $\exp\left( S(E_0) -\beta E_0+ \dots \right)$ where $S(E_0)$ is the CFT density of states at energy $E_0$ and the corrections are sub-leading in the sharpness of $f$ so long as $f$ is roughly constant on scales set by the density of states near $E_0$.

Before proceeding to the main argument, it is useful to remind the reader why the interiors of eternal black holes are of particular concern in the AdS/CFT dictionary. Of course, there are many interesting questions associated with black holes interiors such as those raised in e.g. \cite{Freivogel:2005qh,Balasubramanian:2014hda,Stanford:2014jda,Fischetti:2014uxa,Marolf:2017shp,Marolf:2017vsk}.  But more generally, one might note that whenever there is a clear procedure to construct a bulk state from the vacuum, the corresponding dual QFT state can be obtained by simply applying the same procedure to the dual quantum field theory (QFT). And at a broad conceptual level such representations seem straightforward to find.
In brief, one need only alter the boundary conditions at late times to let certain species of particles escape through infinity, opening what is in effect a decay channel for the bulk system.  One then keeps track of the resulting fluxes through the boundary and waits for the bulk to decay to its ground state.  The fluxes through the AdS boundary can then be recast as sources that transform the original state into the vacuum. Reversing the flow of time then yields the desired construction; technical details are reviewed in appendix A for the interested reader.

However, the status of the supposed above decay is very different in bulks with and without eternal black holes.  When there is no (future) event horizon in the bulk (i.e., no black hole), it is generally clear that the decay can be described by semi-classical bulk physics.  The dictionary for such states is then knowable to reasonable accuracy, though computing it for any given case may require significant effort.  And in cases with a future horizon but no past event horizon one can simply time-reverse the above argument.  On the other hand, in the presence of eternal black holes with both past and future horizons, assuming that the bulk decays into outgoing particles in either the far future or the far past raises the well-known black hole information problem (see e.g. \cite{Harlow:2014yka,Marolf:2017jkr,Unruh:2017uaw} for recent reviews).  In particular, since at present there is no known way to calculate the detailed properties of the flux through the boundary, we cannot use this method to construct the desired dictionary. Instead, the state of the art is to rely on Maldacena's argument \cite{Maldacena:2001kr} at leading order in $G$, and then to add subleading corrections as in
\cite{Jafferis:2015del,Dong:2016eik,Faulkner:2017vdd}. Relying on bulk saddles that dominate path integrals like those in \cite{Maldacena:2001kr} then leads to speculations that non-perturbative constraints may prohibit bulk solutions with certain black hole interiors (e.g., those with complicated topologies \cite{Marolf:2017shp,Marolf:2017vsk,Fu:2018kcp}) from describing states in any complete quantum theory, or at least in theories dual to local QFTs.

The point of this work is to find QFT duals for new bulk geometries by applying arguments like those in \cite{Maldacena:2001kr} to path integrals associated with microcanonical TFD states of the form \eqref{eq:MCTFD}.   The result is closely related to the Lorentz-signature microcanonical path integral introduced by Brown and York in 1992 \cite{Brown:1992bq} which fixes the full energy density and momentum density on each leaf of a foliation of the boundary by spacelike slices. Stationary points of such path integrals are also stationary points of more standard canonical-type path integrals with fixed boundary metric, but the condition for dominance is now maximizing an entropy at fixed energy.  As a result, such path integrals can bring to the fore saddles that fail to dominate in more familiar contexts. We make use of this here to add new entries to the AdS/CFT dictionary involving the interior geometries of small black holes.

We focus below on introducing our microcanonical path integrals and showing that standard two-sided black holes with Killing horizons and Einstein-Rosen-like almost-traversable wormholes are dual to the above QFT states so long as the black holes dominate the microcanonical entropy in the bulk.   Section \ref{sec:MCPI} describes the construction, shows that the above bulk black holes provide stationary points, and demonstrates that -- so long as the saddles are invariant under Euclidean time translations -- the dominant stationary point is the solution of Bekenstein-Hawking highest entropy.
It also discusses generalizations that replace the TFD in \eqref{eq:MCTFD} with states defined by more complicated QFT path integrals.  To be explicit and transparent, we take $f$ in \eqref{eq:MCTFD} to be Gaussian, though the same argument applies readily to any sharply-peaked function $f$.
Section \ref{sec:MCPhys} then shows that the Ryu-Takayanagi (RT) \cite{Ryu:2006bv,Ryu:2006ef} and Hubeny-Ranagamani-Takayanagi (HRT) \cite{Hubeny:2007xt} formulae hold in semiclassical micorcanoical bulk states.  As a consequence, even dropping the above assumption of time-translation-invariant saddles, for states of the form \eqref{eq:MCTFD} the dominant saddle is determined by maximizing the HRT entropy on either boundary.
This section also discusses the extent the semi-classical limit fails when $f$  is taken to be too sharply peaked.   We close with brief comments emphasizing future directions in section \ref{sec:disc}.

\section{Microcanonical Path Integrals}
\label{sec:MCPI}

The semiclassical approximation will be an important tool in analyzing our path integrals below.
As a result, before embarking on our bulk discussion it will be useful to take a quick look at how saddle-point methods reproduce the obvious physics of the microcanonical QFT state using the representation \eqref{eq:MCTFD}.  Here and in section \ref{sec:bulk} below, in order to be explicit we choose $f$ to be Gaussian with
\begin{equation}
\label{eq:Gaussianf}
f(E) = \exp\left(-(E-E_0)^2/4\sigma^2 \right).
\end{equation}

For this $f$, the norm of the state \eqref{eq:MCTFD} is
\begin{equation}
\label{eq:1stnorm}
Z = \langle \psi |\psi \rangle = \int dE e^{-\beta E} e^{S(E)} e^{-(E-E_0)^2/2\sigma^2},
\end{equation}
where $e^{S(E)}$ is the QFT density of states.
Evaluating this integral by the stationary phase method and introducing $S'(E) = \frac{dS}{dE}$, to leading order we find
\begin{equation}
\label{eq:1ststatphase}
-\beta + S'(E) - (E-E_0)/\sigma^2 =0,
\end{equation}
which for small $\sigma$ admits a solution of the form
\begin{equation}
\label{eq:Esolve}
E = E_0 + \sigma^2(S'(E_0) -\beta) + O(\sigma^4).
\end{equation}
In a holographic setting with $E_0$ and $S'(E_0)$ of orders $1/G$ and $1$ in terms of the bulk Newton constant $G$, the second term in \eqref{eq:Esolve} can be neglected as $G \rightarrow 0$ when $\sigma$ grows more slowly than $G^{-1/2}$ or like $G^{-1/2}$ with a suitably small coefficient.  In this case the state $|\psi \rangle$ is indeed dominated by its projection onto states with energy near $E_0$ and we find $\ln Z = S(E_0)-\beta E_0 + O(\sigma^2) +\dots$, where the dots represent higher order corrections to the stationary phase approximation.  The exponent in \eqref{eq:1stnorm} makes clear that any saddles at energies far from $E_0$ are sub-dominant at small $\sigma^2$.  Although the thermodynamic temperature $1/S'(E)$ at $E_0$ generally differs from the arbitrary parameter $\beta$ used in \eqref{eq:MCTFD}, the stationary phase condition \eqref{eq:1ststatphase} is nevertheless satisfied due to the finite value of $\frac{E-E_0}{\sigma^2}$ in the limit $\sigma^2 \rightarrow 0$.

\subsection{Bulk microcanonical path integral and dominant saddles}
\label{sec:bulk}

As in \cite{Maldacena:2001kr}, to translate the state \eqref{eq:MCTFD} into bulk language it is useful to first rewrite \eqref{eq:MCTFD} using path integrals.  This is straightforward if we think of $|\psi\rangle$ as having been obtained from the thermofield double state $|TFD\rangle = \sum_E e^{-\beta E/2} |E\rangle |E\rangle$ by acting with the operator $f(H_L)$ given by \eqref{eq:Gaussianf} with the argument $E$ replaced by the (say) left-Hamiltonian $H_L$. We may then substitute the familiar representation of $|TFD\rangle$ as a path integral over a `cylinder' $C(\beta/2) := X \times [0, \beta/2]$, where we have taken each QFT to live on the spacetime $X \times {\mathbb R}$.  Writing $f(H_L)$ as the Fourier transform
\begin{equation}
\label{eq:FourierF}
f(H_L) = \exp\left(-(H_L-E_0)^2/4\sigma^2 \right) =
\frac{\sigma}{\sqrt{\pi}} \int dt \ e^{it(E_0-H_L)} e^{-\sigma^2 t^2},
\end{equation}
and noting that $e^{-iH_Lt} |TFD\rangle$ is just the time-evolution of the thermo-field double by $t$ on the left, we find
\begin{equation}
\label{eq:MCPIQFT}
|\psi \rangle = f(H_L) |TFD\rangle = \frac{\sigma}{\sqrt{\pi}} \int dt \ e^{itE_0} e^{-\sigma^2 t^2}\int_{{\rm cyl \  of \ length} \tau = \beta/2 +it}{\cal D}\phi_{QFT} \  e^{-I_{QFT}}.
\end{equation}
Here $\int {\cal D}\phi_{QFT}$ denotes the path integral over QFT fields on the given (complex) cylinder ${\cal C}(\tau/2)$ with length $\tau =\beta+2it$ and $I_{QFT}$ is the Euclidean QFT action (evaluated on this complex manifold).  In passing from \eqref{eq:MCTFD} to \eqref{eq:MCPIQFT} we have traded the action of $f(H_L)$ for an integral over an extra parameter $t$ associated with time-evolution of the thermo-field double.  In particular, we have written $|\psi\rangle$ as a superposition of states defined by path integrals.  This will be useful in finding a geometric bulk dual below.

Strictly speaking, we should describe the path integral with fixed QFT fields $\phi_L,\phi_R$ on the two boundaries as giving the components $\langle \phi_L,\phi_R | \psi \rangle$ of $|\psi\rangle $ in $\phi_L,\phi_R$ eigenstates.  However, the more abstract notation in \eqref{eq:MCPIQFT} is useful in translating the expression to the bulk.  Indeed, the dual bulk state is simply given by \eqref{eq:MCPIQFT} with the path integral now being over all bulk geometries whose conformal boundaries match ${\cal C}(\tau)$, and with $I_{QFT}$ replaced by the bulk action $I_{bulk}$.  Again, this path-integral expression should really be taken to compute components $\langle \phi_\Sigma, \Sigma | \psi\rangle$ of $|\psi \rangle$ in a basis defined by fixing appropriate bulk fields $\phi_\Sigma$ on a bulk surface $\Sigma$ that completes ${\cal C}(\tau/2)$ to a closed manifold; i.e., for which $\partial \Sigma = \partial {\cal C}(\tau/2)$ so that $\Sigma \cup {\cal C}(\tau/2)$ gives the full boundary conditions for the bulk path integral.

We wish to show that $|\psi\rangle$ is well-described by a single bulk geometry in the bulk semiclassical limit $G\rightarrow 0$.  This can be done by finding the pair $(t^*,\phi^*_{bulk})$ consisting of a parameter value $t=t_*$ and a bulk field configuration $\phi^*_{bulk}$ that together dominate the bulk version of \eqref{eq:MCPIQFT} in the stationary phase approximation.  Now, it is important here that we find parameters that dominate the {\it state} in the sense that they give the dominant contribution to the norm.  This is different from finding parameters that dominate some particular component $\langle \phi_\Sigma, \Sigma |\psi \rangle$, and the latter will generally depend on $\phi_\Sigma, \Sigma$.  We thus write

\begin{equation}
\label{eq:MCPInorm}
Z : = \langle \psi |\psi \rangle = \frac{\sigma}{\sqrt{2\pi}} \int dt \int_{\partial M ={\cal T}(\tau)} {\cal D} \phi_{bulk} \ e^{-\hat I_{bulk}},
\end{equation}
with the bulk fields $\phi_{bulk}$ satisfying (Euclidean) Aymptotically Locally Anti-de Sitter boundary conditions on a manifold $M$ with conformal boundary $\partial M = T(\tau)$ given by the `torus' $T(\tau) := X \times S^1$ for an $S^1$ of circumference $\tau =\beta +it$ with
\begin{equation}
\label{eq:Ihat}
\hat I_{bulk} : = -itE_0 + \sigma^2 t^2/2 + I_{bulk}
\end{equation}
in terms of the standard bulk action $I_{bulk}$. Note that using \eqref{eq:MCPIQFT} for both $|\psi\rangle$ and $\langle \psi |$ in \eqref{eq:MCPInorm} introduces two separate Fourier transform parameters $t_\psi, t_{\psi^\dagger}$ and that in \eqref{eq:Ihat} we have set $t = t_\psi + t_{\psi^\dagger}$.  This explains the consistency of $\tau = \beta + it$ here with the expression $\tau = \beta +2it$ used in \eqref{eq:MCPIQFT}.

With this understanding, the stationary phase conditions are
\begin{eqnarray}
0&=&\frac{\partial \hat I_{bulk}}{\partial t}\big|_{\phi_{bulk}} = - iE_0 +\sigma^2 t + \frac{\partial I_{bulk}}{\partial t}\big|_{\phi_{bulk}}, \label{eq:vart} \cr
0&=& \frac{\delta \hat I_{bulk}}{\delta \phi_{bulk}}\big|_{t} = \frac{\delta I_{bulk}}{\delta \phi_{bulk}}\big|_{t}. \label{eq:onshell}
\end{eqnarray}
If there is more than one solution to \eqref{eq:onshell} through which the contour of integration can be deformed, the result is dominated by the allowed saddle with the least value for the real part of $\hat I_{bulk}$.

The second condition in \eqref{eq:onshell} requires that the saddle satisfy the usual bulk equations of motion. Solving this condition first, $I_{bulk}$ becomes the on-shell Euclidean bulk action. Noting that the Lorentz-signature bulk action is $iI_{bulk}$ and using Hamilton-Jacobi theory then tells us that
$\frac{\partial I_{bulk}}{\partial t}\big|_{\phi_{bulk}} = iE$ in terms of the Lorentz-signature energy $E$.  The first condition in \eqref{eq:onshell} then yields
\begin{equation}
\label{eq:tstar}
t^* = \frac{i(E_0-E)}{\sigma^2}.
\end{equation}

As already noted on the QFT side, one expects $t_*$ to remain finite as $\sigma^2 \rightarrow 0$ so that $E\rightarrow E_0$.  One can argue this much as in \eqref{eq:Gaussianf}-\eqref{eq:Esolve} by solving the above equations in a slightly different order.  To begin, before imposing the bulk equations of motion let us {\it define} $E$ to be $E: = -i\frac{\partial I_{bulk}}{\partial t}\big|_{\phi_{bulk}}$ and take this to be one of the coordinates on the manifold of bulk fields $\phi_{bulk}$.  The first line of \eqref{eq:vart} then imposes $t=t^*$ as given by \eqref{eq:tstar}.  Furthermore, motivated by the fact that on solutions with with a $U(1)$ Killing field we have $I_{bulk} = (\beta +it)E - S$  in terms of the energy $E$ and entropy $S$, let us more generally define ${\cal S}:= (\beta +it)E  -I_{bulk}$.  Then \eqref{eq:Ihat} can be written
\begin{equation}
\label{eq:imposetstar}
\hat I_{bulk}\big|_{t=t^*} = \beta E-{\cal S}  + (E-E_0)^2/2\sigma^2,
\end{equation}
and one of the conditions from the second line of \eqref{eq:onshell} becomes
\begin{equation}
0 = \frac{\partial \hat I_{bulk}\big|_{t=t^*}}{\partial E} =\beta - \frac{\partial {\cal S}}{\partial E} + (E-E_0)/\sigma^2.
\label{eq:findE}
\end{equation}
In parallel with \eqref{eq:Esolve}, for small $\sigma^2$ \eqref{eq:findE} admits a solution of the form
\begin{equation}
E =E_0 + \sigma^2 \left( \frac{\partial {\cal S}}{\partial E}\big|_{E=E_0} -\beta \right)+ O(\sigma^4).
\end{equation}
Thus as $\sigma^2 \rightarrow 0$ we impose $E=E_0$ and find from \eqref{eq:tstar} that
\begin{equation}
t^* \rightarrow -i  \left( \frac{\partial {\cal S}}{\partial E}\big|_{E=E_0} -\beta \right)
\end{equation}
so $t_*$ remains finite as claimed.  This yields
\begin{equation}
\label{eq:Ivalue}
\hat I_{bulk}\big|_{t^*,\phi_{bulk}^*} = \beta E_0 -{\cal S} + O(\sigma^2).
\end{equation}

If there is more than one allowed solution, the dominant one must minimize \eqref{eq:Ivalue}.
Imposing the rest of the equations of motion, we again note for $U(1)$ symmetric Euclidean black hole solutions that ${\cal S} = (\beta +it)E -I_{bulk}$ becomes the Bekenstein-Hawking entropy $S_{BH}$.  So since $\beta, E_0$ are given constants, if there is more than one solution with $E=E_0$ the maximum entropy solution dominates as expected.  At fixed $\sigma^2$, it also is possible that \eqref{eq:findE} admits other solutions far from $E_0$, but \eqref{eq:imposetstar} shows that any new such solutions will be subdominant at small enough $\sigma^2$.

We may now summarize our results as follows:  In the limit $\sigma^2 \rightarrow 0$, the stationary point dominating the norm \eqref{eq:MCPInorm} is a solution of the bulk equations of motion with boundary torus $T(\tau)$ for $\tau =\beta +it^*$ with the (possibly complex) value of $t^*$ chosen to give the solution energy $E_0$.  Furthermore, if there is more than one such stationary point, since $\beta$ and $E_0$ are the same for all solutions the dominant solution is the one with the largest value of ${\cal S}:= (\beta +it^*)E  -I_{bulk}$.  Assuming the dominant solution to have a $U(1)$ time-translation symmetry this ${\cal S}$ is just the usual Bekenstein-Hawking entropy $S_{BH}$ defined by the action $I_{bulk}$.  More generally, we will show in section \ref{sec:RTHRT} that ${\cal S}$ is the HRT entropy of either boundary.

A priori, the above saddle points are complex as they involve the modulus $\tau = \beta +i t^*$.  However, interesting saddles are provided by {\it real} Euclidean static black holes.  Such black holes have real energies $E$ and we choose $E_0$ real, so from \eqref{eq:tstar} we see that $t^*$ is purely imaginary.  Indeed, as seen from \eqref{eq:findE}, as  $\sigma \rightarrow 0$ the quantity $it$ is precisely the amount by which we must correct the naive Euclidean period $\beta$ of our solution to obtain the physically correct temperature for the given black hole at energy $E_0$.

Such saddles are especially easy to interpret.  Just as for standard Euclidean path integrals, the Euclidean solution determines the leading semi-classical contribution to all bulk correlators in the state $|\psi\rangle$, and the analytic-continuation to Lorentz-signature is the bulk dual to $|\psi\rangle$ to leading order in the bulk semi-classical limit.  In the context of AdS black holes for $E_0$ below that Hawking-Page transition, the end result at this order is that the dominant saddle for our $|\psi\rangle$ will be the saddle point of the standard Euclidean path integral with $X \times S^1$ boundary (without constraints on the length of the $S^1$) that maximizes $S_{BH}$ subject to the constraint $E=E_0$; i.e., it is the two-sided version of the entropically-dominant $U(1)$-symmetric black hole.

\subsection{More general boundaries}

As mentioned in section \ref{sec:Introduction}, the original arguments of \cite{Maldacena:2001kr} are easily generalized to the larger class of states defined by replacing the cylinder ${\cal C}(\beta/2)$ in \eqref{eq:MCPIQFT} with a more general manifold $Y$, perhaps including non-geometric sources for the QFT in addition to the source associated with the metric on $Y$.  In much the same way one can obtain analogous fixed-energy states by acting with $f(H)$ on some boundary of $Y$ and proceeding as above.  However, since $Y$ will generally lack time-translation invariance, fixing the energy on one boundary of $Y$ will not be equivalent to fixing the energy on any other boundary.  Indeed, one may independently choose whether or not to fix the energy on each boundary of $Y$.   For that matter, one may also choose to cut open the manifold $Y$ and to insert an additional $f(H)$ on the cut, fixing the analogue of energy on that surface as well.

For any choices of energies to hold fixed, the end result is much the same as above save for the details of the quantum state and in particular the fact that amplitudes through which microstates contribute to $|\psi\rangle$ no longer take the simple form $e^{-\beta E_0/2}$ as $\sigma^2 \rightarrow 0$.  Indeed, in general there may be no natural analogue of the parameter $\beta$, so it is simplest to leave the analogue of \eqref{eq:Ihat} in the form

\begin{equation}
\label{eq:general}
\hat I_{bulk}\big|_{t^*,\phi_{bulk}^*} = -i\sum_k t_k^*E_{k0} + I_{bulk} + O(\sigma^2),
\end{equation}
where the action is to be computed at the stationary point of $I_{bulk}$ with boundary $\tilde Y$ obtained from $Y$ by inserting appropriate cylinders (with constant sources) of length $it_k$ at each fixed-energy cut or boundary $k$.  The (perhaps complex) parameters $t_k^*$ are chosen so that the bulk has the desired energies $E_{0k}$ on the cuts.  The dominant stationary point is the one minimizing \eqref{eq:general} subject to these constraints, and to leading semi-classical order the norm of the state is \eqref{eq:general}. Since the $E_{k0}$ are fixed, at fixed $t_k^*$ the saddle minimizing the standard action $I_{bulk}$ will dominate.  But the term $-i\sum_k t_k^*E_{k0}$ contributes when comparing saddles with different values of $t_k^*$ and, as in the case with time-translation symmetry described by \eqref{eq:Ivalue}, its role will be to cancel out the additional cost $i\sum_k t_k^*E_{k}$ that would have been assigned by $I_{bulk}$ in deforming $Y$ to $\tilde Y$.

\section{Physics in the Microcanonical Bulk}
\label{sec:MCPhys}

We now provide some short comments on properties of our microcanonical bulk states.  We first show in section \ref{sec:RTHRT} that the RT and HRT relations hold in semiclassical such states.  This is straightforward when the width $\sigma$ if of order $G^{-\alpha}$ for $\alpha \in [-1/2,0)$ as $G\rightarrow 0$.  We then discuss to what extent the semi-classical limit fails for $\sigma$ of order $1$ or smaller as $G \rightarrow 0$.

\subsection{Entanglement in the microcanonical ensemble}
\label{sec:RTHRT}

Our microcanonical path integral is somewhat different from the more familiar fixed-boundary path integrals typically used to address the AdS/CFT correspondence.  One might thus ask if derivations of familiar properties such as the Ryu-Takayangi \cite{Ryu:2006bv,Ryu:2006ef} or Hubeny-Rangamani-Takayanagi \cite{Hubeny:2007xt} formulae continue to hold for bulk states dual to \eqref{eq:MCTFD}.  That this is indeed the case (and similarly for higher-order pertubative corrections in $G$) follows from the fact that the arguments of \cite{Lewkowycz:2013nqa,Faulkner:2013ana,Dong:2016hjy,Dong:2017xht} do not depend on the form of the action away from the conical defect associated with the replica trick since our extra parameter $t$ can be interpreted as merely another such variable on which our actions happens to depend.  More concretely, our on-shell action $\hat I_{bulk}$ differs from the usual action $I_{bulk}$
only by the terms $-itE_0 + \sigma^2 t^2/2$ and in the fact that the manifold on which we evaluate $I_{bulk}$ depends on $t$ as well.    Thus any difference between varying $\hat I_{bulk}$ and $I_{bulk}$ with respect to the replica number must arise through dependence of $t^*$ on $n$.  But since $\hat I_{bulk}$ is stationary under first order variations in $t$ about $t^*|_{n=1}$, we must have $\partial_n \hat I_{bulk} |_{n=1} = \partial_n I_{bulk} |_{n=1}$ and the arguments of
\cite{Lewkowycz:2013nqa,Dong:2016hjy} go through without change, as do the arguments for higher-order corrections in \cite{Faulkner:2013ana,Dong:2016hjy,Dong:2017xht}.  It is only the classical backgrounds and the states of quantum fields on those backgrounds that may differ from those prepared using fixed-boundary path integrals.

Interestingly, this argument shows that our quantity ${\cal S}$ is indeed the entropy so long as we consider a state defined by a boundary manifold with Euclidean time-translation invariance (i.e., by a cylinder ${\cal C}(\beta/2)$) regardless of whether this symmetry is preserved by our bulk saddle.  As noted in section \ref{sec:bulk}, in this case the on-shell action is $\beta E_0 -{\cal S}$.  As usual, taking $n$ replicas effectively multiplies $\beta$ by $n$.  But in the current setting it leaves ${\cal S} = (\beta +it*)E_0 - I_{bulk}$ unchanged since $\beta +it^*$ is fully determined by the condition $E=E_0$ and cannot depend on the value chosen for $\beta$. So denoting the norm of the $n$-fold replica by $Z_n$, one finds as claimed the HRT entropy $S_{HRT} = -n\partial_n[ \ln Z_n - n \ln Z_1]|_{n=1} = -n \partial_n[(1-n)S_{BH}]|_{n=1} = {\cal S}$.

As a result, for Einstein-Hilbert gravity we must from \cite{Lewkowycz:2013nqa,Dong:2016hjy} also have ${\cal S} = A/4$ with $A$ the area of a closed extremal surface in the bulk solution (and presumeably of the smallest such surface).  This in particular helps to justify the natural focus on saddles preserving the Euclidean time-translation invariance of the boundary.  One can find saddles that break this symmetry that correspond to simply displacing a time-translation-symmetric black hole from the origin at $t=0$ and letting the black hole oscillate in Lorentzian time, or equivalently with imaginary Euclidean period\footnote{Alternatively, one can note that the Wick rotation of this solution to Euclidean signature is not periodic but treat it as a limiting case with infinite Euclidean period $\beta +it^*$.  As seen from \eqref{eq:Ivalue}, contributions from such solutions need not be infinitely suppressed.} $i\pi\ell_{AdS} = \beta + it^*$ set by the anti-de Sitter scale $\ell_{AdS}$.  But displacing a black hole of fixed area $A = 4{\cal S}$ in this way increases its energy, and thus ${\cal S} = A/$ must decrease when displaced at fixed energy $E=E_0$.  This makes such solutions subdominant in comparison with those preserving the time-translation symmetry.  Since black holes are dissipative systems, one expects time-independent black holes to similarly dominate over general time-dependent black holes.

\subsection{Finite Width and Bulk Quantum Corrections}

At leading order in the semi-classical approximation, so long as the energy-width $\sigma$ is smaller than a scale of order $G^{-1/2}$, the saddle point selected by our procedure is largely independent of $\sigma$ and has a well-defined limit as $\sigma \rightarrow 0$.  For parametrically large $\sigma$, the window function $f$ is too broad to effect the ensemble and \eqref{eq:MCTFD} effectively reduces to the canonical ensemble, though of course the physics remains semi-classical.

In the full QFT, we can also discuss the $\sigma \rightarrow 0$ limit of \eqref{eq:MCPIQFT}, but the result is quite different.  Indeed, if the QFT lives on $X \times {\mathbb R}$ with compact $X$, the spectrum of $H$ will be discrete.  So unless we choose $E_0$ to agree precisely with one of its eigenvalues, we will find $|\psi \rangle \rightarrow 0$ as $\sigma \rightarrow 0$.  Furthermore, when we do choose $E_0$ to be such an eigenvalue the corresponding eigenstate will generally be unique up to symmetries.  The norm of $|\psi \rangle$ is then set by $e^{-\beta E_0}$ and the dimension of the corresponding symmetry representation and is completely independent of the spacing between energy eigenvalues; i.e., it has nothing to do with what is usually called the density of states $e^S$ at $E_0$.  Indeed, our state then has small entropy in each copy of the QFT.  It must thus be that the HRT relation derived in section \ref{sec:RTHRT} has failed, so such states should certainly not be well-described by a single classical spacetime having a horizon of finite area $4GS$.  Indeed, while the small entanglement might suggest that it would be better described by a disconnected spacetime having no wormhole, we will see below that no single semi-classical geometry can suffice.

The essential point here is described in the appendix of \cite{Marolf:2012xe} (a `state-dependent' interpretation like that advocated in \cite{Papadodimas:2015xma} is not required).  Taking $|\psi \rangle$ to be an exact eigenstate of $H_L$ means that $e^{itH_L}|\psi \rangle =  e^{itE_0}|\psi\rangle$ and (say, left) time-translations leave the state invariant up to an unobservable overall phase.  Thus correlation functions in this state are independent of time, and the probability of e.g. two observers dropped into the wormhole from opposite boundaries to meet in the wormhole interior must be independent of when they are released!  While a semi-classical wormhole has very definite temporal correlations between the two sides, these correlations have been completely smeared out in our eigenstate in accord with standard expectations that energy- and time-resolutions are related by
\begin{equation}
\label{eq:unc}
\Delta t \Delta E \ge \hbar.
\end{equation}
Indeed, starting with e.g. the thermofield-double states and acting with a finite-width $f(H_L)$ as represented above using $e^{itH_L}$ and the fourier transform $\tilde f(t)$, is clear that decreasing the energy-width $\sigma \sim \Delta E$ enlarges the width of temporal correlations in precisely this way.  For $\sigma$ large (say of order $G^{-\alpha}$ for any $\alpha >0$), the effect is small and one may study it in detail by recalling that $|TFD\rangle$ gives the Hartle-Hawking state of quantum fields on the associated eternal black hole backgound and acting with some $f(H_L)$ on the Hartle-Hawking state.  In contrast,  for $\sigma$ of order $G^0$ or smaller the fact that $\Delta t$ is of order $G^0$ or larger means that the smearing in time gives a superposition of distinct semi-classical bulk spacetimes, though one may still quantitatively study the effect on boundary correlators.

Note that, despite the small entanglement at small $\sigma$, the situation is generally not improved by supposing that the bulk might be described by a wormhole-free disconnected geometry, as for small $\sigma$ our state is sharply peaked both with respect to the eigenvalues of $H_L$ and with respect to the eigenvalues of $H_R$.  As a result, a bulk dual given by a single disconnected semi-classical geometry would require separate symmetries under translations by $H_L$ and $H_R$; i.e., each of the two connected components must be time-independent.  While such solutions can exist, too few of them are known to correspond to generic energy eigenvalues $E_0$ (though see e.g. \cite{Bena:2014qxa} for progress in attempts to find sufficiently large sets of such solutions following what a strong version of what is now called the fuzzball proposal of \cite{Lunin:2001jy}).

\section{Discussion}
\label{sec:disc}

In parallel with Maldacena's classic result \cite{Maldacena:2001kr}, the above work used a path integral representation of the microcanonical thermalfield-double state \eqref{eq:MCTFD} at energy $E_0$ to argue that in holographic QFTs this state is dual to the standard Kruskal-like two-sided extension of the AdS black hole (see figure \ref{fig:wormhole}) that dominates the microcanonical ensemble at $E_0$. As shown in section \ref{sec:RTHRT}, in general this dominance is determined by maximizing the HRT entropy of one boundary, though for solutions with a time-translation symmetry this agrees with the usual Bekenstein-Hawking entropy of the bulk.  Here it is of course important that one consider the full bulk theory asymptotic to some AdS$_d \times K$ and not simply the truncation to gravity on AdS$_d$ as at some energies the entropically dominant black hole will be localized on $K$; see e.g. \cite{Horowitz:1999uv}.

Although different from more familiar constructions with fixed boundary-metric, microcanonical path integrals are nevertheless dominated by semi-classical stationary points with familiar properties so long as the width $\sigma$ of the microcanonical ensemble is not too small.  In particular, the RT and HRT relations hold as usual.  However, for small widths $\sigma$ of order $G^0$ or smaller, quantum fluctuations in the bulk prohibit our microcanonical states from being described by a single semi-classical geometry.

Our construction is closely related to the Lorentz-signature microcanonical path integral introduced by Brown and York in 1992 \cite{Brown:1992bq}.  Indeed, the reader should consult e.g. \cite{Brown:1993ke} for explicit examples.  Their construction fixed not only the total energy, but also the energy and momentum densities on some foliation of the boundary with fixed spatial metric.  Much as in our analysis above, one expects to obtain their full formalism by acting with additional window-functions $f$ that enact these further constraints. In doing so, one should not be able to produce an eigenstate of the local energy density operator $\rho = -T_t^t$ as the commutator $[\rho(x),\rho(y)]$ is non-trivial at $x=y$.  As a result, constraining $\rho(x)$ at the semi-classical level will require somewhat-arbitrary choices regarding the final quantum state, though these choices should make little difference at leading semi-classical order where fixing the full distribution is a well-defined operation.  Indeed, a natural choice is to fix the energy distribution only on length-scales longer than some $\lambda$ with fluctuations shorter than $\lambda$ giving an entropy-like contribution to the norm of the state.  See e.g. \cite{Horowitz:1996th,Horowitz:1996cj} for an explicit discussion of the associated ensembles in a black string context, though literature on fluid approximations to microscopic theories will contain many other references as well.

It would be interesting to construct such fixed energy-and-momentum-density path integrals in detail by introducing the above constraints.  As noted in \cite{Brown:1992bq}, the associated bulk dynamics satisfy the standard equations of motion but with non-standard boundary conditions on the time-time and time-space components of the bulk metric that fix the associated parts of the boundary stress tensor instead of the metric on the boundary.  Such boundary conditions were termed `Neumann' in \cite{Compere:2008us}, and for each component they are the metric analogue of the `alternate' boundary conditions discussed for scalars in \cite{Klebanov:1999tb}. Imposing such boundary conditions on all components of the bulk metric would lead to a dual CFT operator violating unitarity bounds, and so presumably also to bulk ghosts as in \cite{Andrade:2011dg}.  It would thus be useful to carefully analyze the AdS-analogue of the boundary conditions in \cite{Brown:1992bq} for the presence or absence of ghosts.

Many other points of potential interest clearly remain to be explored as well. One is to give a detailed account of quantum teleportation in microcanonical TFDs in analogy with the canonical treatment of \cite{Maldacena:2017axo}.  Given the similarities between the entanglements in microcanonical and canonical such states, one would expect their analysis to go through unchanged so long as our width $\sigma$ is sufficiently large, and perhaps even more generally so long as one allows sufficiently long time for the teleportation to occur in relation to the temporal width determined by \eqref{eq:unc}.

Another outstanding issue is to determine how generally one can use our technique to focus on bulk saddles that are subdominant in path integrals with fixed boundary metrics.  For example, can one use such techniques to find QFT duals of arbitrary (say, time-symmetric) Euclidean bulk solutions by taking the associated energies large enough to disfavor certain boundary cycles from pinching off in the bulk?  Such pinch-offs generally require the spacetime in such regions to resemble empty global AdS and thus to have small entropy relative to black holes at the same energy.  If successful, this would eliminate the concerns of \cite{Marolf:2017shp} and establish that the holographic entropy cone \cite{Bao:2015bfa} does not depend on e.g. needing to consider bulk duals with arbitrary numbers of fermion fields.  We expect that using fixed-energy techniques to focus on otherwise-subdominant saddles will provide useful in analyzing other facets of bulk entanglement as well.

\section*{Acknowledgements}
It is a pleasure to thank Xi Dong, Dan Harlow, Veronika Hubeny, Henry Maxfield, Jonathan Oppenheim, Mukund Rangamani, Douglas Stanford, Aron Wall, and Jason Wien for useful conversations. I also thank  Juan Pedraza,  Andrew Svesko, Watse Sybesma, and Manus Visser for pointing out a typo in an earlier version (including the published version). This work was supported in part by a U.S. National Science Foundation under grant number PHY15-04541 and by the University of California. Much of this work was performed at the Aspen Center for Physics during the 2017 ACP workshop Information in Quantum Field Theory, which was supported by National Science Foundation grant PHY-1607611.

\appendix

\section{Building horizon-free bulk states from the vacuum}

So long as the the dual QFT flows to a CFT$_d$ UV fixed point having a least one operator
${\cal O}$ for which both the dimension $\Delta$ of ${\cal O}$ and the dimension $d-\Delta$ of the conjugate source both satisfy the corresponding unitarity bound, we can turn on a relevant local coupling between our system and an auxiliary one QFT$_{aux}$ with a much larger density of states.  Choosing an appropriate time-dependent coupling that turns on at $t=0$, we can imagine starting the joint system in the product state $|\psi \rangle_{bulk} \otimes |0\rangle_{aux}$ involving the desired bulk state $|\psi\rangle_{bulk}$ and the vacuum $|0 \rangle_{aux}$ of the auxiliary system, and evolving to the far future.  Here the notation $|\psi\rangle_{bulk}$ emphasizes that we begin with a bulk description of the state for which dual QFT description may not yet be known. Under any such coupling one generically expects the systems to equilibrate and, in the limit of large density of states QFT$_{aux}$ to do so with exponentially small energy remaining in the original QFT.

In particular, for large enough QFT$_{aux}$ we can approximate the equilibrium state at some far future time $t_f$ as $|\Psi_{eq}(t_f) \rangle \approxeq |0\rangle \times |\psi \rangle_{aux}$ in terms of the bulk vacuum $|0\rangle$ and some pure state $|\psi \rangle_{aux}$ of QFT$_{aux}$.  So long as the above bulk dynamics are well-described by effective field theory, this $\psi \rangle_{aux}$ can in principle be calculated without using the dual QFT description of the bulk.

To recover a dual QFT description $|\psi \rangle_{QFT}$ of the original state $|\psi \rangle_{bulk}$, we now simply reverse the procedure replacing the bulk with the dual QFT.   In particular, starting with $|0\rangle \times |\psi \rangle_{aux}$  at time $t_f$ (where $|0\rangle$ is now the vacuum of the dual QFT) and evolving the coupled system backwards in time to $t=0$ using the dual QFT dynamics yields a state well-approximated by
$|\psi \rangle_{QFT} \otimes |0\rangle_{aux}$.  Repeating this construction with larger and larger auxiliary systems and taking a limit then gives $|\psi \rangle_{QFT}$ exactly, or at least to the extent that the full dynamics is in fact captured by bulk effective field theory.

\bibliographystyle{jhep}

\begin{thebibliography}{99}

\bibitem{Jafferis:2015del}
D.~L. Jafferis, A.~Lewkowycz, J.~Maldacena and S.~J. Suh, \emph{{Relative
  entropy equals bulk relative entropy}},
  \href{http://dx.doi.org/10.1007/JHEP06(2016)004}{\emph{JHEP} {\bfseries 06}
  (2016) 004}, [\href{https://arxiv.org/abs/1512.06431}{{\ttfamily
  1512.06431}}].

\bibitem{Botta-Cantcheff:2015sav}
M.~Botta-Cantcheff, P.~Martínez and G.~A. Silva, \emph{{On excited states in
  real-time AdS/CFT}},
  \href{http://dx.doi.org/10.1007/JHEP02(2016)171}{\emph{JHEP} {\bfseries 02}
  (2016) 171}, [\href{https://arxiv.org/abs/1512.07850}{{\ttfamily
  1512.07850}}].

\bibitem{Dong:2016eik}
X.~Dong, D.~Harlow and A.~C. Wall, \emph{{Reconstruction of Bulk Operators
  within the Entanglement Wedge in Gauge-Gravity Duality}},
  \href{http://dx.doi.org/10.1103/PhysRevLett.117.021601}{\emph{Phys. Rev.
  Lett.} {\bfseries 117} (2016) 021601},
  [\href{https://arxiv.org/abs/1601.05416}{{\ttfamily 1601.05416}}].

\bibitem{Christodoulou:2016nej}
A.~Christodoulou and K.~Skenderis, \emph{{Holographic Construction of Excited
  CFT States}}, \href{http://dx.doi.org/10.1007/JHEP04(2016)096}{\emph{JHEP}
  {\bfseries 04} (2016) 096},
  [\href{https://arxiv.org/abs/1602.02039}{{\ttfamily 1602.02039}}].

\bibitem{Faulkner:2017vdd}
T.~Faulkner and A.~Lewkowycz, \emph{{Bulk locality from modular flow}},
  \href{http://dx.doi.org/10.1007/JHEP07(2017)151}{\emph{JHEP} {\bfseries 07}
  (2017) 151}, [\href{https://arxiv.org/abs/1704.05464}{{\ttfamily
  1704.05464}}].

\bibitem{Marolf:2017kvq}
D.~Marolf, O.~Parrikar, C.~Rabideau, A.~Izadi~Rad and M.~Van~Raamsdonk,
  \emph{{From Euclidean Sources to Lorentzian Spacetimes in Holographic
  Conformal Field Theories}},
  \href{http://dx.doi.org/10.1007/JHEP06(2018)077}{\emph{JHEP} {\bfseries 06}
  (2018) 077}, [\href{https://arxiv.org/abs/1709.10101}{{\ttfamily
  1709.10101}}].

\bibitem{Maldacena:2001kr}
J.~M. Maldacena, \emph{{Eternal black holes in anti-de Sitter}},
  \href{http://dx.doi.org/10.1088/1126-6708/2003/04/021}{\emph{JHEP} {\bfseries
  04} (2003) 021}, [\href{https://arxiv.org/abs/hep-th/0106112}{{\ttfamily
  hep-th/0106112}}].

\bibitem{Skenderis:2009ju}
K.~Skenderis and B.~C. van Rees, \emph{{Holography and wormholes in 2+1
  dimensions}},
  \href{http://dx.doi.org/10.1007/s00220-010-1163-z}{\emph{Commun. Math. Phys.}
  {\bfseries 301} (2011) 583--626},
  [\href{https://arxiv.org/abs/0912.2090}{{\ttfamily 0912.2090}}].

\bibitem{Balasubramanian:2014hda}
V.~Balasubramanian, P.~Hayden, A.~Maloney, D.~Marolf and S.~F. Ross,
  \emph{{Multiboundary Wormholes and Holographic Entanglement}},
  \href{http://dx.doi.org/10.1088/0264-9381/31/18/185015}{\emph{Class. Quant.
  Grav.} {\bfseries 31} (2014) 185015},
  [\href{https://arxiv.org/abs/1406.2663}{{\ttfamily 1406.2663}}].

\bibitem{Marolf:2015vma}
D.~Marolf, H.~Maxfield, A.~Peach and S.~F. Ross, \emph{{Hot multiboundary
  wormholes from bipartite entanglement}},
  \href{http://dx.doi.org/10.1088/0264-9381/32/21/215006}{\emph{Class. Quant.
  Grav.} {\bfseries 32} (2015) 215006},
  [\href{https://arxiv.org/abs/1506.04128}{{\ttfamily 1506.04128}}].

\bibitem{Maxfield:2016mwh}
H.~Maxfield, S.~Ross and B.~Way, \emph{{Holographic partition functions and
  phases for higher genus Riemann surfaces}},
  \href{http://dx.doi.org/10.1088/0264-9381/33/12/125018}{\emph{Class. Quant.
  Grav.} {\bfseries 33} (2016) 125018},
  [\href{https://arxiv.org/abs/1601.00980}{{\ttfamily 1601.00980}}].

\bibitem{Marolf:2017shp}
D.~Marolf, M.~Rota and J.~Wien, \emph{{Handlebody phases and the polyhedrality
  of the holographic entropy cone}},
  \href{http://dx.doi.org/10.1007/JHEP10(2017)069}{\emph{JHEP} {\bfseries 10}
  (2017) 069}, [\href{https://arxiv.org/abs/1705.10736}{{\ttfamily
  1705.10736}}].

\bibitem{Marolf:2017vsk}
D.~Marolf and J.~Wien, \emph{{The Torus Operator in Holography}},
  \href{http://dx.doi.org/10.1007/JHEP01(2018)105}{\emph{JHEP} {\bfseries 01}
  (2018) 105}, [\href{https://arxiv.org/abs/1708.03048}{{\ttfamily
  1708.03048}}].

\bibitem{Fu:2018kcp}
Z.~Fu, A.~Maloney, D.~Marolf, H.~Maxfield and Z.~Wang, \emph{{Holographic
  complexity is nonlocal}},
  \href{http://dx.doi.org/10.1007/JHEP02(2018)072}{\emph{JHEP} {\bfseries 02}
  (2018) 072}, [\href{https://arxiv.org/abs/1801.01137}{{\ttfamily
  1801.01137}}].

\bibitem{Hawking:1982dh}
S.~W. Hawking and D.~N. Page, \emph{{Thermodynamics of Black Holes in anti-De
  Sitter Space}}, \href{http://dx.doi.org/10.1007/BF01208266}{\emph{Commun.
  Math. Phys.} {\bfseries 87} (1983) 577}.

\bibitem{Asplund:2008xd}
C.~T. Asplund and D.~Berenstein, \emph{{Small AdS black holes from SYM}},
  \href{http://dx.doi.org/10.1016/j.physletb.2009.02.043}{\emph{Phys. Lett.}
  {\bfseries B673} (2009) 264--267},
  [\href{https://arxiv.org/abs/0809.0712}{{\ttfamily 0809.0712}}].

\bibitem{Hanada:2016pwv}
M.~Hanada and J.~Maltz, \emph{{A proposal of the gauge theory description of
  the small Schwarzschild black hole in AdS$_5\times$S$^5$}},
  \href{http://dx.doi.org/10.1007/JHEP02(2017)012}{\emph{JHEP} {\bfseries 02}
  (2017) 012}, [\href{https://arxiv.org/abs/1608.03276}{{\ttfamily
  1608.03276}}].

\bibitem{Yaffe:2017axl}
L.~G. Yaffe, \emph{{Large $N$ phase transitions and the fate of small
  Schwarzschild-AdS black holes}},
  \href{http://dx.doi.org/10.1103/PhysRevD.97.026010}{\emph{Phys. Rev.}
  {\bfseries D97} (2018) 026010},
  [\href{https://arxiv.org/abs/1710.06455}{{\ttfamily 1710.06455}}].

\bibitem{Berenstein:2018lrm}
D.~Berenstein, \emph{{Submatrix deconfinement and small black holes in AdS}},
  \href{https://arxiv.org/abs/1806.05729}{{\ttfamily 1806.05729}}.

\bibitem{Fidkowski:2003nf}
L.~Fidkowski, V.~Hubeny, M.~Kleban and S.~Shenker, \emph{{The Black hole
  singularity in AdS / CFT}},
  \href{http://dx.doi.org/10.1088/1126-6708/2004/02/014}{\emph{JHEP} {\bfseries
  02} (2004) 014}, [\href{https://arxiv.org/abs/hep-th/0306170}{{\ttfamily
  hep-th/0306170}}].

\bibitem{Klosch:1995qv}
T.~Klosch and T.~Strobl, \emph{{Classical and quantum gravity in
  (1+1)-dimensions. Part 2: The Universal coverings}},
  \href{http://dx.doi.org/10.1088/0264-9381/13/9/007}{\emph{Class. Quant.
  Grav.} {\bfseries 13} (1996) 2395--2422},
  [\href{https://arxiv.org/abs/gr-qc/9511081}{{\ttfamily gr-qc/9511081}}].

\bibitem{Gao:2016bin}
P.~Gao, D.~L. Jafferis and A.~Wall, \emph{{Traversable Wormholes via a Double
  Trace Deformation}},
  \href{http://dx.doi.org/10.1007/JHEP12(2017)151}{\emph{JHEP} {\bfseries 12}
  (2017) 151}, [\href{https://arxiv.org/abs/1608.05687}{{\ttfamily
  1608.05687}}].

\bibitem{Freivogel:2005qh}
B.~Freivogel, V.~E. Hubeny, A.~Maloney, R.~C. Myers, M.~Rangamani and
  S.~Shenker, \emph{{Inflation in AdS/CFT}},
  \href{http://dx.doi.org/10.1088/1126-6708/2006/03/007}{\emph{JHEP} {\bfseries
  03} (2006) 007}, [\href{https://arxiv.org/abs/hep-th/0510046}{{\ttfamily
  hep-th/0510046}}].

\bibitem{Stanford:2014jda}
D.~Stanford and L.~Susskind, \emph{{Complexity and Shock Wave Geometries}},
  \href{http://dx.doi.org/10.1103/PhysRevD.90.126007}{\emph{Phys. Rev.}
  {\bfseries D90} (2014) 126007},
  [\href{https://arxiv.org/abs/1406.2678}{{\ttfamily 1406.2678}}].

\bibitem{Fischetti:2014uxa}
S.~Fischetti, D.~Marolf and A.~C. Wall, \emph{{A paucity of bulk entangling
  surfaces: AdS wormholes with de Sitter interiors}},
  \href{http://dx.doi.org/10.1088/0264-9381/32/6/065011}{\emph{Class. Quant.
  Grav.} {\bfseries 32} (2015) 065011},
  [\href{https://arxiv.org/abs/1409.6754}{{\ttfamily 1409.6754}}].

\bibitem{Harlow:2014yka}
D.~Harlow, \emph{{Jerusalem Lectures on Black Holes and Quantum Information}},
  \href{http://dx.doi.org/10.1103/RevModPhys.88.015002}{\emph{Rev. Mod. Phys.}
  {\bfseries 88} (2016) 015002},
  [\href{https://arxiv.org/abs/1409.1231}{{\ttfamily 1409.1231}}].

\bibitem{Marolf:2017jkr}
D.~Marolf, \emph{{The Black Hole information problem: past, present, and
  future}}, \href{http://dx.doi.org/10.1088/1361-6633/aa77cc}{\emph{Rept. Prog.
  Phys.} {\bfseries 80} (2017) 092001},
  [\href{https://arxiv.org/abs/1703.02143}{{\ttfamily 1703.02143}}].

\bibitem{Unruh:2017uaw}
W.~G. Unruh and R.~M. Wald, \emph{{Information Loss}},
  \href{http://dx.doi.org/10.1088/1361-6633/aa778e}{\emph{Rept. Prog. Phys.}
  {\bfseries 80} (2017) 092002},
  [\href{https://arxiv.org/abs/1703.02140}{{\ttfamily 1703.02140}}].

\bibitem{Brown:1992bq}
J.~D. Brown and J.~W. York, Jr., \emph{{The Microcanonical functional integral.
  1. The Gravitational field}},
  \href{http://dx.doi.org/10.1103/PhysRevD.47.1420}{\emph{Phys. Rev.}
  {\bfseries D47} (1993) 1420--1431},
  [\href{https://arxiv.org/abs/gr-qc/9209014}{{\ttfamily gr-qc/9209014}}].

\bibitem{Ryu:2006bv}
S.~Ryu and T.~Takayanagi, \emph{{Holographic derivation of entanglement entropy
  from AdS/CFT}},
  \href{http://dx.doi.org/10.1103/PhysRevLett.96.181602}{\emph{Phys. Rev.
  Lett.} {\bfseries 96} (2006) 181602},
  [\href{https://arxiv.org/abs/hep-th/0603001}{{\ttfamily hep-th/0603001}}].

\bibitem{Ryu:2006ef}
S.~Ryu and T.~Takayanagi, \emph{{Aspects of Holographic Entanglement Entropy}},
  \href{http://dx.doi.org/10.1088/1126-6708/2006/08/045}{\emph{JHEP} {\bfseries
  08} (2006) 045}, [\href{https://arxiv.org/abs/hep-th/0605073}{{\ttfamily
  hep-th/0605073}}].

\bibitem{Hubeny:2007xt}
V.~E. Hubeny, M.~Rangamani and T.~Takayanagi, \emph{{A Covariant holographic
  entanglement entropy proposal}},
  \href{http://dx.doi.org/10.1088/1126-6708/2007/07/062}{\emph{JHEP} {\bfseries
  07} (2007) 062}, [\href{https://arxiv.org/abs/0705.0016}{{\ttfamily
  0705.0016}}].

\bibitem{Lewkowycz:2013nqa}
A.~Lewkowycz and J.~Maldacena, \emph{{Generalized gravitational entropy}},
  \href{http://dx.doi.org/10.1007/JHEP08(2013)090}{\emph{JHEP} {\bfseries 08}
  (2013) 090}, [\href{https://arxiv.org/abs/1304.4926}{{\ttfamily 1304.4926}}].

\bibitem{Faulkner:2013ana}
T.~Faulkner, A.~Lewkowycz and J.~Maldacena, \emph{{Quantum corrections to
  holographic entanglement entropy}},
  \href{http://dx.doi.org/10.1007/JHEP11(2013)074}{\emph{JHEP} {\bfseries 11}
  (2013) 074}, [\href{https://arxiv.org/abs/1307.2892}{{\ttfamily 1307.2892}}].

\bibitem{Dong:2016hjy}
X.~Dong, A.~Lewkowycz and M.~Rangamani, \emph{{Deriving covariant holographic
  entanglement}}, \href{http://dx.doi.org/10.1007/JHEP11(2016)028}{\emph{JHEP}
  {\bfseries 11} (2016) 028},
  [\href{https://arxiv.org/abs/1607.07506}{{\ttfamily 1607.07506}}].

\bibitem{Dong:2017xht}
X.~Dong and A.~Lewkowycz, \emph{{Entropy, Extremality, Euclidean Variations,
  and the Equations of Motion}},
  \href{http://dx.doi.org/10.1007/JHEP01(2018)081}{\emph{JHEP} {\bfseries 01}
  (2018) 081}, [\href{https://arxiv.org/abs/1705.08453}{{\ttfamily
  1705.08453}}].

\bibitem{Marolf:2012xe}
D.~Marolf and A.~C. Wall, \emph{{Eternal Black Holes and Superselection in
  AdS/CFT}},
  \href{http://dx.doi.org/10.1088/0264-9381/30/2/025001}{\emph{Class. Quant.
  Grav.} {\bfseries 30} (2013) 025001},
  [\href{https://arxiv.org/abs/1210.3590}{{\ttfamily 1210.3590}}].

\bibitem{Papadodimas:2015xma}
K.~Papadodimas and S.~Raju, \emph{{Local Operators in the Eternal Black Hole}},
  \href{http://dx.doi.org/10.1103/PhysRevLett.115.211601}{\emph{Phys. Rev.
  Lett.} {\bfseries 115} (2015) 211601},
  [\href{https://arxiv.org/abs/1502.06692}{{\ttfamily 1502.06692}}].

\bibitem{Bena:2014qxa}
I.~Bena, M.~Shigemori and N.~P. Warner, \emph{{Black-Hole Entropy from
  Supergravity Superstrata States}},
  \href{http://dx.doi.org/10.1007/JHEP10(2014)140}{\emph{JHEP} {\bfseries 10}
  (2014) 140}, [\href{https://arxiv.org/abs/1406.4506}{{\ttfamily 1406.4506}}].

\bibitem{Lunin:2001jy}
O.~Lunin and S.~D. Mathur, \emph{{AdS / CFT duality and the black hole
  information paradox}},
  \href{http://dx.doi.org/10.1016/S0550-3213(01)00620-4}{\emph{Nucl. Phys.}
  {\bfseries B623} (2002) 342--394},
  [\href{https://arxiv.org/abs/hep-th/0109154}{{\ttfamily hep-th/0109154}}].

\bibitem{Horowitz:1999uv}
G.~T. Horowitz, \emph{{Comments on black holes in string theory}},
  \href{http://dx.doi.org/10.1088/0264-9381/17/5/320}{\emph{Class. Quant.
  Grav.} {\bfseries 17} (2000) 1107--1116},
  [\href{https://arxiv.org/abs/hep-th/9910082}{{\ttfamily hep-th/9910082}}].

\bibitem{Brown:1993ke}
J.~D. Brown and J.~W. York, Jr., \emph{{Microcanonical action and the entropy
  of a rotating black hole}},
  \href{https://arxiv.org/abs/gr-qc/9303012}{{\ttfamily gr-qc/9303012}}.

\bibitem{Horowitz:1996th}
G.~T. Horowitz and D.~Marolf, \emph{{Counting states of black strings with
  traveling waves}},
  \href{http://dx.doi.org/10.1103/PhysRevD.55.835}{\emph{Phys. Rev.} {\bfseries
  D55} (1997) 835--845},
  [\href{https://arxiv.org/abs/hep-th/9605224}{{\ttfamily hep-th/9605224}}].

\bibitem{Horowitz:1996cj}
G.~T. Horowitz and D.~Marolf, \emph{{Counting states of black strings with
  traveling waves. 2.}},
  \href{http://dx.doi.org/10.1103/PhysRevD.55.846}{\emph{Phys. Rev.} {\bfseries
  D55} (1997) 846--852},
  [\href{https://arxiv.org/abs/hep-th/9606113}{{\ttfamily hep-th/9606113}}].

\bibitem{Compere:2008us}
G.~Compere and D.~Marolf, \emph{{Setting the boundary free in AdS/CFT}},
  \href{http://dx.doi.org/10.1088/0264-9381/25/19/195014}{\emph{Class. Quant.
  Grav.} {\bfseries 25} (2008) 195014},
  [\href{https://arxiv.org/abs/0805.1902}{{\ttfamily 0805.1902}}].

\bibitem{Klebanov:1999tb}
I.~R. Klebanov and E.~Witten, \emph{{AdS / CFT correspondence and symmetry
  breaking}},
  \href{http://dx.doi.org/10.1016/S0550-3213(99)00387-9}{\emph{Nucl. Phys.}
  {\bfseries B556} (1999) 89--114},
  [\href{https://arxiv.org/abs/hep-th/9905104}{{\ttfamily hep-th/9905104}}].

\bibitem{Andrade:2011dg}
T.~Andrade and D.~Marolf, \emph{{AdS/CFT beyond the unitarity bound}},
  \href{http://dx.doi.org/10.1007/JHEP01(2012)049}{\emph{JHEP} {\bfseries 01}
  (2012) 049}, [\href{https://arxiv.org/abs/1105.6337}{{\ttfamily 1105.6337}}].

\bibitem{Maldacena:2017axo}
J.~Maldacena, D.~Stanford and Z.~Yang, \emph{{Diving into traversable
  wormholes}}, \href{http://dx.doi.org/10.1002/prop.201700034}{\emph{Fortsch.
  Phys.} {\bfseries 65} (2017) 1700034},
  [\href{https://arxiv.org/abs/1704.05333}{{\ttfamily 1704.05333}}].

\bibitem{Bao:2015bfa}
N.~Bao, S.~Nezami, H.~Ooguri, B.~Stoica, J.~Sully and M.~Walter, \emph{{The
  Holographic Entropy Cone}},
  \href{http://dx.doi.org/10.1007/JHEP09(2015)130}{\emph{JHEP} {\bfseries 09}
  (2015) 130}, [\href{https://arxiv.org/abs/1505.07839}{{\ttfamily
  1505.07839}}].

\end{thebibliography}
	\cleardoublepage
\phantomsection
\renewcommand*{\bibname}{References}

\end{document}